Article Reentry Shuttle after Cath 6 30 06

# A New Method of Atmospheric Reentry for Space Ships*

**Alexander Bolonkin**
C&R, 1310 Avenue R, #F-6, Brooklyn, NY 11229, USA
T/F 718-339-4563, aBolonkin@juno.com , aBolonkin@gmail.com, http://Bolonkin.narod.ru

## Abstract

In recent years, industry has produced high-temperature fiber and whiskers. The author examined the atmospheric reentry of the USA's Space Shuttles and proposed the use of high temperature tolerant parachute for atmospheric air braking. Though it is not large, a light parachute decreases Shuttle speed from 8 km/s to 1 km/s and Shuttle heat flow by 3-4 times. The parachute surface is opened with backside so that it can emit the heat radiation efficiently to Earth-atmosphere. The temperature of parachute is about 1000-1300° C. The carbon fiber is able to keep its functionality up to a temperature of 1500-2000° C. There is no conceivable problem to manufacture the parachute from carbon fiber. The proposed new method of braking may be applied to the old Space Shuttles as well as to newer spacecraft designs.

## Introduction

In 1969 Alexander A. Bolonkin applied a new method of global optimization to the problem of atmospheric reentry of spaceships [1 p. 188]. The general analysis presented an additional method to the well-known method of outer space to Earth-atmosphere reentry ("high-speed corridor"). There is a low-speed corridor when the total heat is less than in a conventional high-speed passage. In that time for significantly decreasing the speed of a spaceship retro- and landing rocket engine needed to be used. That requires a lot of fuel. It is not acceptable for modern spaceships. Nowadays the textile industry produces heat resistant fiber that can be used for a new parachute system to be used in a high-temperature environment.

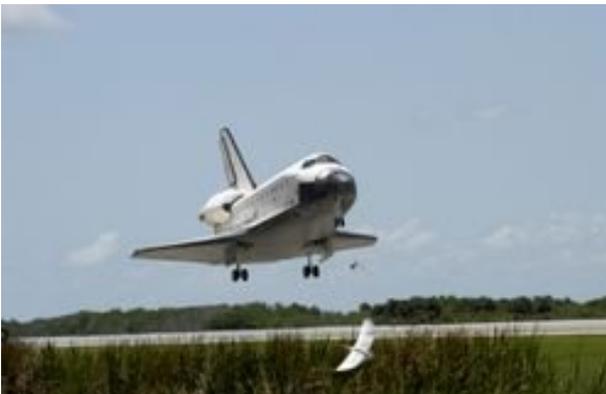 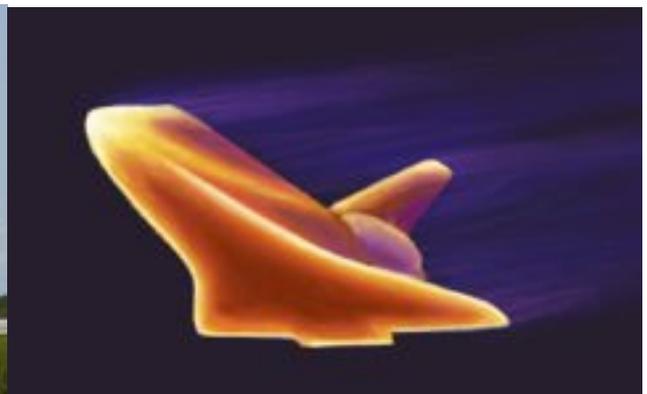

**Fig.1.** Space Shuttle "Atlantic".　　**Fig.2.** The outside of the Shuttle heats to over 1,550 °C during reentry.

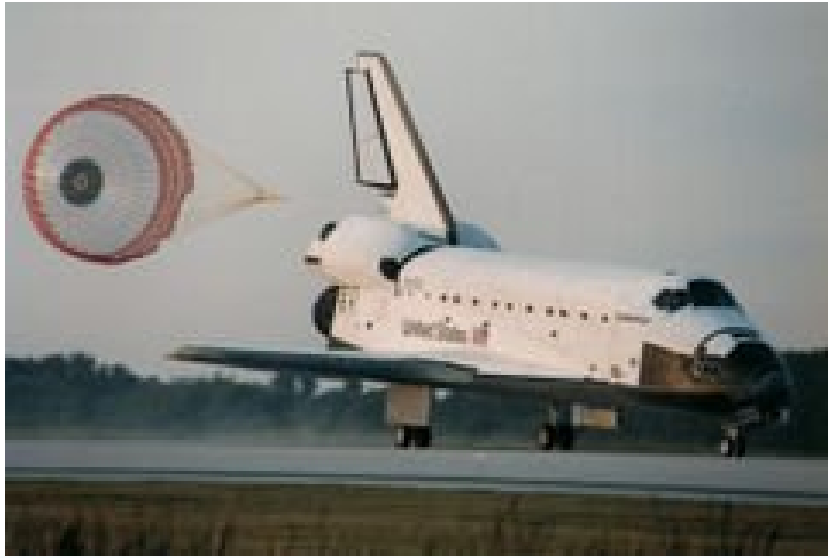

**Fig.3.** Endeavour deploys drag chute after touch-down.

# Theory

1. Equations of spaceship reentry are:

$$\dot{r} = \frac{R_0}{R} V \cos\theta,$$

$$\dot{H} = V \sin\theta,$$

$$\dot{V} = -\frac{D + D_P}{m} - g \sin\theta,$$   (1)

$$\dot{\theta} = \frac{L + L_P}{mV} - \frac{g}{V}\cos\theta + \frac{V\cos\theta}{R} + 2\omega_E \cos\varphi_E,$$

where $r$ is range of ship flight, m; $R_0 = 6,378,000$ is radius of Earth, m; $R$ is radius of ship flight from Earth's center, m; $V$ is ship speed, m/s; $H$ is ship altitude, m; $\theta$ is trajectory angle, radians; $D$ is ship drag, N; $D_P$ is parachute drag, N; $m$ is ship mass, kg; $g$ is gravity at altitude $H$, m/s²; $L$ is ship lift force, N; $L_P$ is parachute lift force, N; $\omega_E$ is angle Earth speed; $\varphi_E = 0$ is lesser angle between perpendicular to flight plate and Earth polar axis; $t$ is flight time, sec.

The magnitudes in equations (1) compute as:

$$g = g_0\left(\frac{R_0}{R_0 + H}\right)^2, \quad \rho = a_1 e^{(H-10000)/b}, \quad a_1 = 0.414, \quad b = 6719,$$

$$Q = \frac{0.5 \cdot 11040 \cdot 10^4}{R_n^{0.5}} \left(\frac{\rho}{\rho_{SL}}\right)^{0.5} \left(\frac{V}{V_{CO}}\right)^{3.15}, \quad R_n = \sqrt{\frac{S_P}{\pi}},$$   (2)

$$T_1 = 100\left(\frac{Q}{\varepsilon C_S} + \left(\frac{T_2}{100}\right)^4\right)^{1/4}, \quad T = T_1 - 273,$$

$$D_P = 0.5 C_{DP}\rho V S_P, \quad L_P = D_P / 3, \quad L = 2\alpha\rho V S, \quad D = L / 4,$$

where: $g_0 = 9.81$ m/s² is gravity at Earth surface; $\rho$ is air density, kg/m³; $Q$ is heat flow in 1 m²/s of parachute, J/s·m²; $R_n$ (or $R_p$) is parachute radius, m; $S_P$ (or $S_m$) is parachute area, m²; $\rho_{SL} = 1.225$ kg/m³ is air density at sea level; $V_{CO} = 7950$ m/s is circle orbit speed; $T_1$ is temperature of parachute in stagnation point in Kelvin, °K; $T$ is temperature of parachute in stagnation point in centigrade, °C; $T_2$ is temperature of the standard atmosphere at given altitude, °K; $D_P$ is parachute drag, N; $L_P$ is parachute lift force, N (the ram-air parachute can produce lift force up 1/3 from its drag); $D$ is ship drag, N; $L$ is ship lift force, N; $C_{DP} = 1$ is parachute drag coefficient; $a = 295$ m/s is sound speed; $\alpha = 40^o = 0.7$ rad is ship attack angle.

The control is following: if $d\theta/dt > 0$ the all lift force $L = L_P = 0$. When the Shuttle riches the low speed the parachute area can be decreased or parachute can be detached. That case not computed. Used control is not optimal.

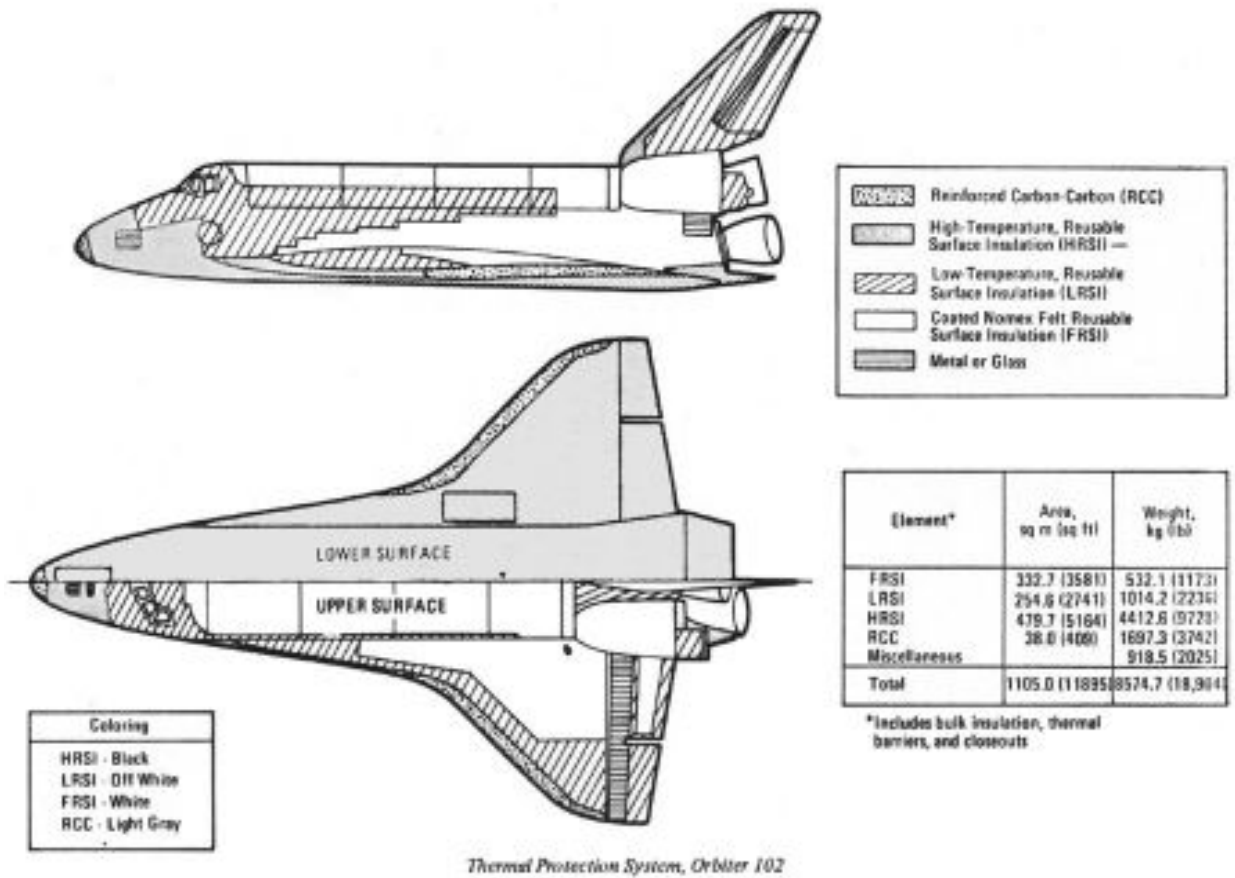

**Fig. 4.** Space Shuttle Thermal Protection System Constituent Materials

The results of integration are presented below. Used data: parachute area are $S_P = 1000, 2000, 4000$ m$^2$ ($R_P = 17.8, 25.2, 35.7$ m); $m = 104,000$ kg. The dash line is data of the Space Shuttle without a parachute,

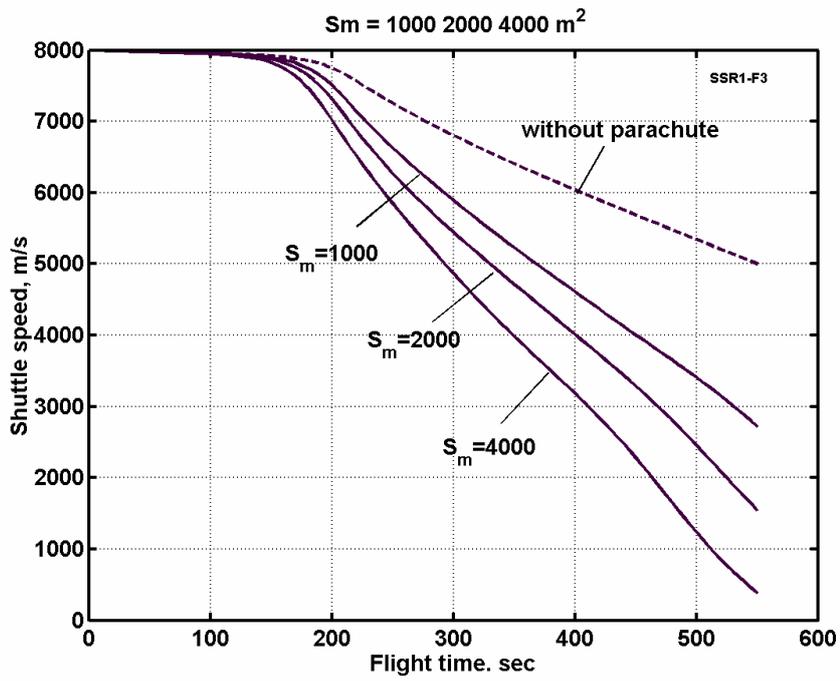

**Fig. 5.** Decreasing of Space Shuttle speed with parachute and without it. $S_\mathrm{m} = S_\mathrm{P}$.

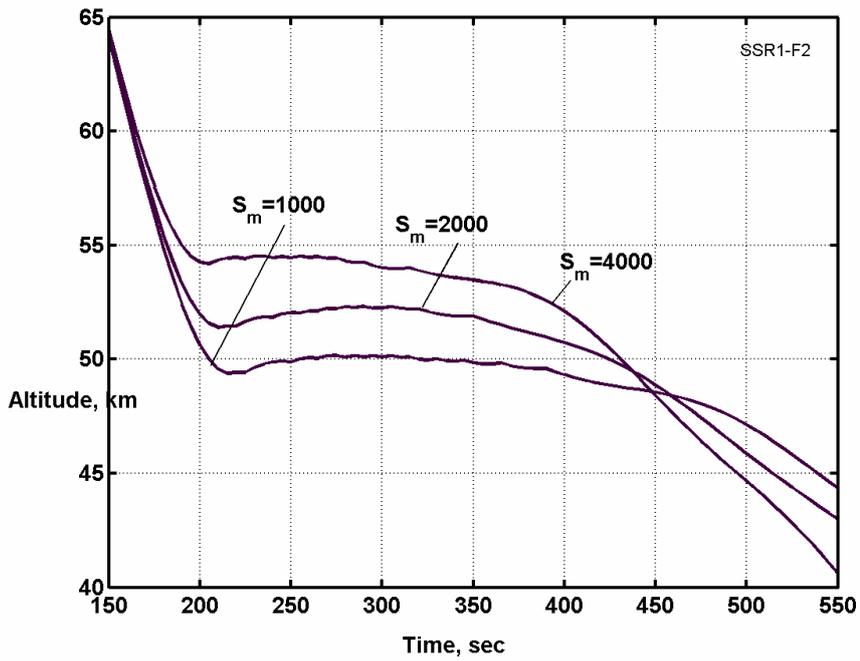

**Fig. 6.** Space Shuttle trajectory with parachute.

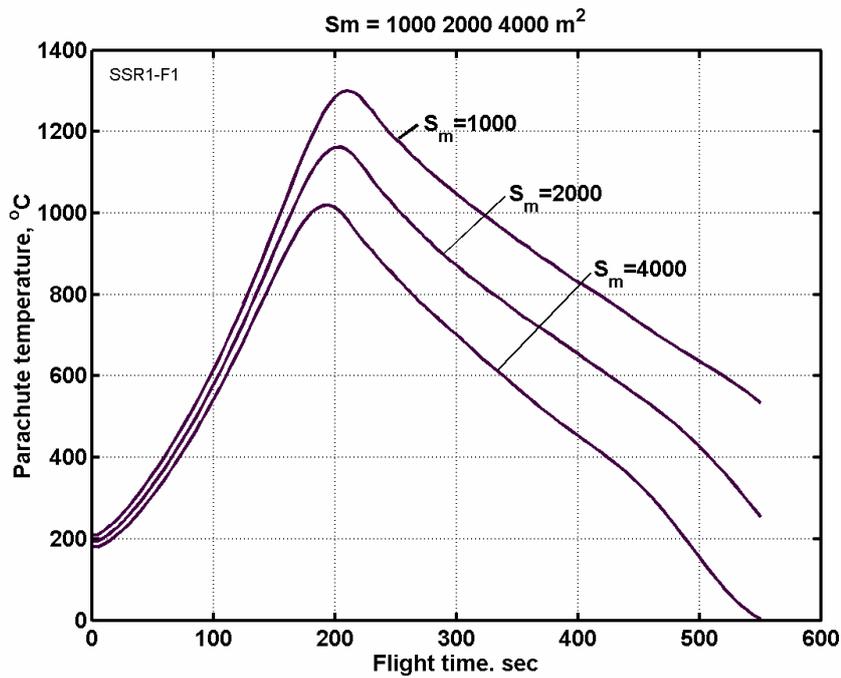

**Fig. 7.** Temperature of parachute at stagnation point.

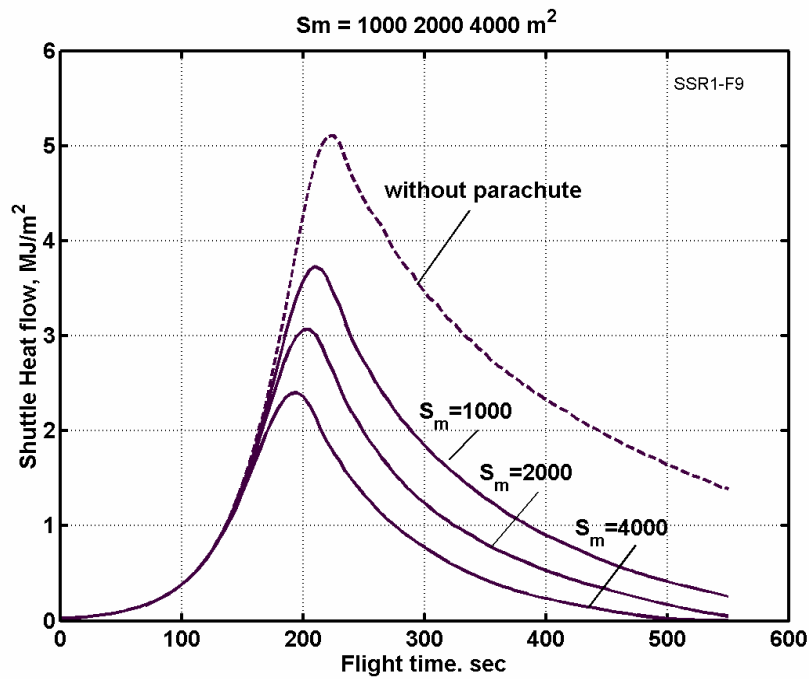

**Fig. 8.** Heat flow through 1 m²/s of Space Shuttle surface at stagnation point with parachute and without it.

## Discussion of results

Fig.5 shows the parachute significantly decreases the shuttle speed from 8000 m/s to 350 - 2900 m/s after 550 sec of reentry flight. Practically, the Space Shuttle overpasses the heat barrier (maximum of heat flow) near 200 sec into its reentry (see Fig. 8). The heat flow depends on the power 3.15 from speed (see the second equation in (2)) and the speed strongly influences the heat flow. For example, the decreasing of speed in two times decreases the heat flows in 8.9 times!

Fig. 6 shows: at altitude 41 - 44 km the ship has speed 350 - 2900 m/s which is acceptable for high speed vehicle in short time of reentry.

Fig. 7 shows the maximum temperature in a stagnation point of the parachute. It is 1000 - 1300$^{\circ}$C. The parachute can be made from carbon fiber that can keep the temperature 1500 - 2000$^{\circ}$C (carbon melting temperature is over 3000$^{\circ}$C). At present a carbon fiber composite matters uses by Shuttle for leader edges of Shuttle where temperature reaches 1550$^{\circ}$C.

Fig. 8 gives the heat flow through 1 m$^2$/s of Shuttle without or with a parachute. If we continue flight time up 750 sec, the special total heat will be following: without parachute it is $11.84 \times 10^8$ J/m$^2$, for parachute having area 1000 m$^2$ - $7 \times 10^8$ J/m$^2$, for 2000 m$^2$ parachute - $5 \times 10^8$ J/m$^2$, for 4000 m$^2$ parachute - $3.5 \times 10^8$ J/m$^2$. That is about 1.7 - 3.4 times less then without parachute. It means the future Space Shuttles can have a different system of heat protection and a modern design can be made lighter and cheaper.

## Estimation Parachute System

The weight of the parachute system and a comparison with current heat protection is key moment for this innovative method. Industry has produced many metal and mineral fibers and whiskers having very high tensile stress at high temperatures. Let us to estimate the mass of parachute system. Assume the carbon fiber having the maximum tensile stress $\sigma = 565$ kg/mm$^2$ ($\sigma = 5.65 \times 10^9$ N/m$^2$) at temperature $T = 1500$ - 2000$^{\circ}$C is used for parachute. Let us take the safety margin 2.3 - 3. That means $\sigma = 150$ kg/mm$^2$ for canopy and $\sigma = 200$ kg/mm$^2$ for cord. The fiber density is taken $\gamma = 3000$ kg/m$^3$.

The computation is presented in Table #1.

**Table #1**. Parachute data.

| | | | |
|---|---|---|---|
| Parachute area $S_p = S_m$, m$^2$ | 1000 | 2000 | 4000 |
| Reference parachute radius $R_p$, m | 17.8 | 25.2 | 35.7 |
| Max. parachute pressure $P_p$, N/m$^2$ | 1250 | 2000 | 6000 |
| Parachute surface $S_{pc} = 2\pi R_p^2$ m$^2$ | 2000 | 4000 | 8000 |
| Parachute thickness $\delta = P_p R_p / 2\sigma$, mm | 0.0074 | 0.0076 | 0.0072 |
| Mass of canopy $M_c = S_{pc}\delta\gamma$, kg | 45 | 90 | 171 |
| Mass of cord, kg | 66 | 132 | 258 |
| Total mass, kg | 111 | 226 | 429 |
| Max. brake force, kN | 1250 | 1800 | 2400 |
| Add. Max. overload, g | 1.25 | 1.8 | 2.4 |

Currently, the mass of the heat protection in Shuttle is 9575 kg. If we decrease this protection proportional the decreasing of the heat flow (in 2 - 3 times) we save the 4 - 6 tons of Shuttle mass.

At the present time, the changing of hundreds of hull protection tiles after every flight takes two weeks and very costly to do. The new method requires only a few tile replacements (maximum temperature is less) or allows using a protective cooling method.

  The Command Module of spacecraft "Apollo" had a heat protection of approximately 1/3 of the total take-off/touchdown weight. The gain to be had from a new method reentering may be significantly more.

### Conclusion

The widespread production of high temperature fibers and whiskers allows us to design high-temperature tolerant parachutes, which may be used by space apparatus of all types for braking in a rarified planet atmosphere. The parachute has open backside surface that rapidly emits the heat radiation to outer space thereby quickly decreasing the parachute temperature. The proposed new method significantly decreases the maximum temperature and heat flow to main space apparatus. That decreases the heat protection mass and increases the useful load of the spacecraft. The method may be also used during an emergency reentering when spaceship heat protection is damaged (as in horrific instance of the Space Shuttle "Columbia").